\RequirePackage{fix-cm}
\documentclass[twocolumn,epjc3]{svjour3}

\smartqed  

\usepackage{graphicx}
\usepackage{dcolumn}
\usepackage{bm}
\usepackage{hyperref}
\usepackage{textcomp}
\usepackage{amsmath}
\usepackage{amssymb}
\usepackage{cuted}

\journalname{Eur. Phys. J. C}

\begin{document}

\title{Probing the QCD critical point through linearized sound wave dispersion and attenuation at finite baryon density}

\author{Abhishek Paul\thanksref{e1,addr1}
        \and
        Mehedi Kalam\thanksref{e2,addr2} 
        \and
        Prabir Kumar Haldar\thanksref{e3,addr1}
}

\thankstext{e1}{e-mail: paulabhishek.res@gmail.com}
\thankstext{e2}{e-mail: mehedikalam.phys@aliah.ac.in}
\thankstext{e3}{e-mail: prabirkrhaldar@gmail.com (corresponding author)}

\institute{Department of Physics, Cooch Behar Panchanan Barma University, Vivekananda Street, Cooch Behar, West Bengal -736101, India \label{addr1}
           \and
           Department of Physics, Aliah University, New Town, Kolkata 700160, West Bengal, India \label{addr2}
}

\date{Received: date / Accepted: date}

\maketitle

\begin{abstract}
We investigate the linearized sound-channel response of a finite baryon density Israel-Stewart hydrodynamic system coupled to a parameterized equation of state containing a critical end point. The attenuation of sound is characterized by a threshold wavelength below which sound perturbations are strongly damped. We systematically study the dependence of this threshold scale on the non-universal parameters of the Beam Energy Scan Theory (BEST) equation of state framework. In our present study, the threshold wavelength remains finite but develops a pronounced peak in the vicinity of the critical end point, indicating strong absorption of sound modes on fireball-scale wavelengths. We further estimate the impact of this enhanced attenuation on the linearized anisotropic flow response and find a reduction of the elliptic and triangular flow harmonics near the critical region. Our results establish a phenomenological connection between critical sound attenuation, Mach cone like correlations, and the linearized flow response in baryon rich QCD matter. These results suggest a non-monotonic beam-energy dependence of the sound attenuation scale. Together with a corresponding suppression of Mach cone like correlations and the linearized anisotropic flow response, they may provide a consistent signature that the system trajectory passes close to the critical point of the chosen equation of state.
\end{abstract}

\section{Introduction}
One fo the key objective of the relativistic heavy Ion Collision Experiment (RHIC-E) is to constrain the critical end point (CEP) of the QCD matter through the Beam Energy Scan (BES) program \cite{wong1994}. At vanishing baryon density, first principle lattice calculations confirm the presence of a continuous phase transition in the baryon-antibaryon symmetric regime near $T\simeq155\ \text{MeV}$ \cite{aoki2009,borsanyi2010,bazavov2012}. At high baryonic potentials, the QCD phase diagram is expected to contain a first order transition line \cite{Asakawa1989,Halasz1998,Stephanov2004}, terminating at the critical end point (CEP). By varying the collision energy $\sqrt{S_{NN}}$, in the BES program, different sets of temperature $T$ and baryon chemical potential $\mu_{B}$ in the QCD phase diagram can be explored.

Historically, the azimuthal distributions of emitted particles have provided crucial insights into the reaction dynamics of ultra-relativistic nuclear collisions. Phenomenological studies of particle emission patterns \cite{Ghosh2001} and multiplicity fluctuations across different azimuthal bins \cite{Ghosh2004} have demonstrated that initial-state spatial properties strongly dictate final-state distributions. To model this evolution dynamically, relativistic viscous hydrodynamics can provide a useful framework for describing the space-time evolution of the QGP. The thermodynamic and transport properties of such a hydrodynamic frame can be connected to experimentally accessible observables \cite{Heinz2013}. As a long-wavelength effective theory, hydrodynamics is organized as a gradient expansion and it becomes increasingly sensitive to microscopic relaxation scales near a critical point. In the vicinity of the critical end point, QCD is expected to belong to the same universality class as the three-dimensional Ising model \cite{aoki2009}.

The growth of the correlation length near the critical end point modifies relaxation dynamics significantly. At the same time, the dissipative contributions, particularly those associated with the bulk viscosity gets strongly enhanced. These effects are expected to increase the attenuation of sound modes near the CEP.

Now in order to study such effects, an accurate description of QCD is required. The current knowledge of finite density QCD equation of state from first principle is derived by using various extrapolation techniques, such as the Taylor expansion\cite{Gavai2003,Allton2005,Bellwied2015,Borsanyi2010bp,Bazavov2017,Bazavov2020} analytical continuation from imaginary chemical potential \cite{Bellwied2015b,Borsanyi2020,Borsanyi2018} etc. However, these approaches are limited to the small chemical potential ranges.

To extend this information towards the higher baryon density range, which is accessible to the BES-II program, the Beam Energy Scan Theory (BEST) collaboration has developed a parameterized equation of state mapped to a QCD background \cite{Parotto2020,Karthein2021}. The mapping contains nonuniversal parameters that control the size and shape of the critical region, while additional mapping parameters determine its orientation in the $(T,\mu_B)$ plane. This construction provides a controlled framework for studying the sensitivity of thermodynamic and hydrodynamic observables to the assumed critical-region structure. In this work, we investigate sound attenuation along thermodynamic trajectories passing near a CEP of the chosen equation of state, located at, $\left(T_{c},\mu_{c}\right)=\left(143.2 ~\text{MeV},350~\text{MeV}\right)$. 

Within the linearized hydrodynamic theory, the long-wavelength sound dispersion relation takes the generalized form: $\omega(k)=\pm c_{s}k-i\Gamma_{s}k^{2}+\mathcal{{O}}(k^{3})$, where $c_{s}$ is the speed of sound and $\Gamma_{s}$ is the sound attenuation coefficient. Using this relation we have calculated the threshold wavelength $\lambda_{\text{th}}$, below which all the sound perturbations are strongly attenuated.

In idealized hydrodynamic limit, the threshold wavelength $\lambda_{\text{th}}$ diverges near the critical end point \cite{Hasanujjaman2020}. In our present parameterized EoS, we obtained a finite $\lambda_{\text{th}}$, whose magnitude and spatial width depends on the non-universal parameters. When the magnitude of  $\lambda_{\text{th}}$ exceeds the characteristic transverse dimension of the QGP fireball, sound modes with wavelengths relevant to system become strongly attenuated, thereby reducing the ability of the medium to sustain long-wavelength collective sound propagation.

In this present work, we have used causal second order Israel-Stewart hydrodynamics \cite{Israel1979}, which incorporates finite time relaxation equations. Using the attenuation profile in this framework, we have constructed a phenomenological estimation of the linearized anisotropic flow response. In this estimate, the critical modification of sound propagation near the CEP is mapped onto the linear harmonic dependence of the flow response, allowing us to study the associated changes in the linearized elliptical and triangular harmonics. 

The paper is organized as follows: In Section II, we described the construction of the parameterized equation of state and the selection of the non-universal parameters governing the critical-region mapping. In Sec. III, we introduced the finite density I-S framework and derived the linearized dispersion relation. In Sec. IV, we present an analytical expression for the threshold wavelength $\lambda_{\text{th}}$ and calculated the corresponding phenomenological linearized anisotropic flow responses. Sec. V is dedicated to results and discussion.In Section VI, we present the concluding remarks. Throughout this work, we use the mostly plus Minkowski metric, $g^{\mu\nu}=\text{diag}(-,+,+,+)$ and natural units $c=\hbar=k_{B}=1$. 

\section{The Equation of State}
\subsection{Construction of the EoS}

To construct the EoS of the lattice QCD \cite{Parotto2020}, first a suitable parameterization is used to define the universal scaling behavior of the EoS in the 3D Ising model near CEP and then it is mapped onto the QCD via non universal change of variables. Next, the EoS of the Ising model is used to estimate the critical contribution to the expansion coefficients from the lattice QCD. Finally, the full pressure is reconstructed by matching the lattice QCD at $\mu_{B}=0$, including the correct critical behavior.

For the scaling of the EoS of 3D Ising model, near the CEP, the following parameterization for the magnetization $M$, the magnetic field $h$ and the reduced temperature $r=\ (T-T_{C})/T_{C}$ is used \cite{Nonaka:2004kp,Guida:1996ep,Schofield:1969zz,Bluhm:2006CPOD}:

\begin{align}
    M&= M_{0}R^{\beta}\theta,\\
    h&= h_{0} R^{\beta\delta} \tilde{h}(\theta),\\
    r&= R\ (1- \theta^{2}),
\end{align}

where $M_{0}$, $h_{0}$ are the normalization constants, $\beta$ and $\delta$ are the critical Ising exponents and $\tilde{h}(\theta)=\theta (1+a \ \theta^{2}+b \ \beta^{4})$, with $a=-0.76201$ and $b=0.00804$ \cite{Guida:1996ep}. The parameters abides by the following relations, $R\geq0$ and $\lvert\theta\rvert\leq \theta_{0} \simeq 1.154$, where $\theta_{0}$ is the first non-trivial zero of $\tilde{h}(\theta)$.

Following \cite{Nonaka:2004kp,Guida:1996ep}, the 3D Ising model scaling EoS can be written as
\begin{equation}
    P_{Ising}(R,\theta)=-h_{0}M_{0}R^{2-\alpha} \left[g(\theta)-\theta\tilde{h}(\theta)\right]
\end{equation}

Now a nontrivial mapping is used to transform the critical thermodynamics between the Ising model variables $(h,r)$ and lattice QCD coordinates $(T,\mu_{B})$. The most generalized transformation between these two requires six parameters,

\begin{align}
    \frac{T-T_{C}}{T_{C}}&=w\left(r\rho\sin\alpha_{1}+h\sin\alpha_{2}\right),\\
    \frac{\mu_{B}-\mu_{BC}}{T_{C}}&=w\left(-r\rho\cos\alpha_{1}-h\cos\alpha_{2}\right),
\end{align}

where $T_{c}$ and $\mu_{Bc}$ are the chosen CEP coordinates in the $T-\mu$ plane; $\alpha_{1}$ and  $\alpha_{2}$ are the respective angles between  $r$ and $h$ axes and the lines of constant $T$. $w$ and $\rho$ represents the global and relative rescaling parameters respectively.

If we consider that the contribution of the Taylor coefficients cab be separated from the contribution of lattice QCD and can be represented as a sum of Ising contribution  due to the critical point and the non-Ising one, then we can write \cite{Parotto:2019qcd}:

\begin{equation}
    T^{4}c_{n}^{LAT}=T^{4}c_{n}^{Non-Ising}(T)+T^{4}_{c}c_{n}^{Ising}(T),
\end{equation}

and the total pressure may be expressed as,
\begin{eqnarray}
    P(T,\mu_{B})&=&T^{4}\sum_{n}c_{n}^{Non-Ising}(T)\left(\frac{\mu_{B}}{T}\right)^{n}\nonumber\\
    &&+T^{4}_{c}P^{Ising}\left(T,\mu_{B}\right).
\end{eqnarray}

From this above equation, all required thermodynamic quantities are extracted.

\subsection{The choice of the EoS parameters}

If the curvature of the crossover line of the transition at $\mu_{B}=0$ is approximated with the shape of parabola \cite{Bazavov2017,Bellwied2015b}, then we can write:

\begin{equation}
    T=T_{0}+\kappa T_{0}\left(\frac{\mu_{B}}{T_{0}}\right)^{2}+\mathcal{O}(\mu_{B}^{4}),
\end{equation}

where $T_{0}=155\text{MeV}$ and $\kappa=-0.0149$ \cite{Bellwied2015b} are the transition temperature and curvature of the transition line at $\mu_{B}=0$ respectively.Due to this approximation, the total number of required parameters comes down from $6$ to $4$.

\begin{equation}
\text{Here}, \alpha_1 = \tan^{-1} \left(2\frac{\kappa}{T_0}\mu_{B,C}\right).
\end{equation}

In this work, we consider the following choices of the four parameters:
\begin{equation*}
    \mu_{Bc} = 350~\text{MeV}, \qquad T_{c}\simeq 143.2~\text{MeV} 
\end{equation*}

The global rescaling parameter is varied within the range
\begin{equation*}
    1\leq w\leq2,
\end{equation*}

while the relative rescaling parameter is varied within
\begin{equation*}
    1\leq\rho\leq3.
\end{equation*}

For the baseline parameter set, we have used
\begin{equation*}
    \alpha_1=3.85^\circ, \qquad \alpha_2=93.85^\circ.
\end{equation*}

\section{The hydrodynamic description of the fluid and propagation of the perturbation}

We are using Israel-Stewart(I-S) hydrodynamics to study the evolution of the QGP \cite{Israel1979}. Due to strong interactions, the net baryon number remains conserved throughout the QGP evolution. The energy-momentum tensor in I-S hydrodynamics is given by
\begin{equation}
    T^{\mu\nu}= \mathcal{E} u^{\mu}u^{\nu}+P \Delta^{\mu \nu}+ 2 h^{(\mu} u^{\nu)}+\tau^{\mu\nu},
\end{equation}
where $u^{\mu}$ is the four velocity vector subjected to the normalization condition, $u_{\mu}u^{\mu}=-1$, $P$ being the thermodynamic pressure, $\tau^{\mu\nu}$ $(=\Pi \Delta^{\mu\nu}+\pi^{\mu\nu})$ is the dissipative viscous stress. $\Pi$ is the bulk viscous pressure, $\Delta^{\mu\nu}$ is the projection operator which satisfies, $u_{\mu}\Delta^{\mu\nu}=u_{\nu}\Delta{\mu\nu}=0$. $\pi^{\mu\nu}$ is the shear viscous stress that obeys, $\pi^{\mu}_{\mu}=h^{\mu}u_{\mu}=\tau^{\mu\nu}u_{\mu}=0$ and the symmetric tensor $h^{(\mu} u^{\nu)}=\frac{1}{2}\left(h^{\mu}u^{\nu}+h^{\nu}u^{\mu}\right) $.

The particle four flow vector is defined as, 
\begin{equation}
    N^{\mu}=nu^{\mu}+n^{\mu},
\end{equation}
where $n$ is the baryon number density and $n^{\mu}$ is the particle diffusion current.
The heat flux four vector is defined as
\begin{equation}
    q^{\mu}=h^{\mu}-n^{\mu}(\mathcal{E}+P).
\end{equation}

Here we are choosing Landau-Lifshitz (LL) frame \cite{Landau1987} to study a system having non-zero net baryon number density.In LL frame, $h^{\mu}=0$ and thus $n^{\mu}=-nq^{\mu}/(\mathcal{E} + P)$. So the energy-momentum tensor takes the following form
\begin{equation}
    T^{\mu\nu}=\mathcal{E} u^{\mu}u^{\nu}+\left(P+\Pi\right)\Delta^{\mu\nu}+\pi^{\mu\nu}
\end{equation}

 The different viscous fluxes are given by \cite{Israel1979},

\begin{align}
    \Pi &= -\frac{1}{3} \zeta \left[\partial_{\mu}u^{\mu}+\beta_{0}D\Pi_{\alpha\beta}-\alpha_{1}\partial_{\mu}q^{\mu}\right], \label{eqn15}\\
    \pi^{\mu\nu} &= -2 \eta \Delta^{\mu\nu\alpha\beta}\left[\partial_{\alpha}u_{\beta}+\beta_{2}D\pi_{\alpha\beta}-\alpha_{1}\partial_{\alpha}q_{\beta}\right], \label{eqn16}\\
    q^{\mu} &= \kappa T \Delta^{\mu\nu} \left[ \frac{nT}{(\mathcal{E}+P)}\left(\partial_{\nu}\alpha\right)-\beta_{1}Dq_{\nu} \right. \nonumber \\
    &\quad \left. +\alpha_{0}\partial_{\nu}\Pi+\alpha_{1}\partial_{\lambda}\Pi^{\lambda}_{\nu} \right],\label{eqn17}
\end{align}
where, $D\equiv u^{\mu}\partial_{\mu}\ \text{and} \  \alpha=\mu/T$, is known as the thermal potential. $\eta, \zeta \ \text{and} \ \kappa$ are the coefficients of shear viscosity, bulk viscosity and thermal conductivity respectively. $\beta_{0}, \beta_{1} \ \text{and} \ \beta_{2}$ are the relaxation coefficients, $\alpha_{0} \text{and} \alpha_{1}$ are the coupling constants. The relaxation time for the bulk pressure $\left(\tau_{\Pi}\right)$, heat flux $\left(\tau_{q}\right)$ and shear tensor $\left(\tau_{\pi}\right)$ are related ot the relaxation coefficients as \cite{Muronga:2006zx},
\begin{equation}
    \tau_{\Pi}=\zeta \beta_{0},\ \tau_{q}=\kappa T\beta_{1}, \ \tau_{\pi}=2 \eta \beta_{2},
\end{equation}

and the various relaxation lengths are defined as, 
\begin{equation}
    l_{\Pi q}=\zeta \alpha_{0},\ l_{q\Pi}=\kappa T \alpha_{0}, \ l_{q\pi}=\kappa T\alpha_{1}, \ l_{\pi q} = 2 \eta \alpha_{1}
\end{equation}

At ultra-relativistic limit, one can obtain these following relations \cite{Israel1979},
\begin{align}
    \alpha_{0}&=6 \beta^{-2} P^{-1},\\
    \alpha_{1}&= - \frac{1}{4}P^{-1},\\
    \beta_{0}&= 216 \beta^{-4}P^{-1},\\
    \beta_{1} &\simeq \frac{5}{4}P^{-1},\\
    \beta_{2}&\simeq \frac{3}{4} P^{-1},
\end{align}
where $\beta=m/T$. 

To investigate the propagation of small disturbances in the fluid, we perform a linear response analysis around a static, homogeneous and isotropic background in global thermal equilibrium. We introduce a small amplitude longitudinal perturbation equation moving along z axis, proportional to $\sim e^{-i\left(\omega t - k z\right)}$. We express the hydrodynamic variables as a sum of their constant equilibrium background values and a small fluctuations, as $X(t,z)\ = X_{0}+\delta X\left({t,z}\right)$, where $\delta X<<X_{0}$. We can write,

\begin{align}
    \mathcal{E}\left(t,z\right)&= \mathcal{E}_{0}+ \delta \mathcal{E} \left(t,z\right)\\
    n\left(t,z\right)&=n_{0}+\delta n \left(t,z\right)\\
    u^{\mu}\left(t,z\right)&= u^{\mu}_{0}+\delta u^{\mu}\left(t,z\right)
\end{align}

Since the dissipative quantities vanish in a static background  $\left(u^{\mu}_{0}=(1,0,0,0)\right)$, therefore we can write,
\begin{align}
    \Pi &= \delta \Pi\left(t,z\right),\\
    q^{\mu}&= \delta q^{\mu}\left(t,z\right).\\
    \pi^{\mu\nu}&= \delta \pi^{\mu\nu}\left(t,z\right).
\end{align}

 Now the conservation of the energy-momentum tensor $\left(T^{\mu\nu}\right)$ and the net-baryon number $\left(N^{\mu}\right)$ of the fluid is given by

\begin{align}
    \partial_{\mu}T^{\mu\nu}&=0,\label{eqn31}\\
     \text{and} \qquad \partial_{\mu}N^{\mu}&=0.\label{eqn32}
\end{align}

Using equation $(\ref{eqn15}),(\ref{eqn16}),(\ref{eqn17}),(\ref{eqn31}),\ \text{and} \ (\ref{eqn32})$,we can obtain the following linearized equations:
\begin{strip}
\begin{align}
    \partial_0\left(\delta \mathcal{E}\right)+\left(\mathcal{E}_{0}+P_{0}\right)\partial_{i}\left(\delta u^{i}\right)&=0\\
    \left(\mathcal{E}_{0}+P_{0}\right)\partial_{0}\left(\delta u^{z}\right)+\partial_{z}\left(c_{s}^{2}\delta \varepsilon+c_{n}^{2}\partial n+\Pi+\pi^{zz}\right)&=0\\
    \partial \left(\delta n\right)+n_{0}\partial_{i}\left(\delta u^{i}\right)-\frac{n_{0}}{\left(\mathcal{E}_{0}+P_{0}\right)}\partial_{i}q^{i}&=0\\
    \Pi+\frac{1}{3}\zeta\beta_{0}\partial_{0}\Pi+\frac{1}{3}\zeta\partial_{i}\left(\delta u^{i}\right)-\frac{1}{3}\zeta \alpha_{0}\partial_{i}q^{i}&=0\\
    \pi^{ij}+2\eta\left[\partial^{<i}\delta u^{j>}+\beta_{2}\partial_{0}\pi^{ij}-\alpha_{1}\partial^{<i}q^{j>}\right]&=0\\
    q^{i}+\kappa T_{0}\beta_{1}\partial_{0}q^{i}-\left(\frac{\partial \alpha}{\partial \mathcal{E}}\right)_{n}\frac{n_0T_0}{\left(\mathcal{E}_0+P_0\right)}\kappa T_{0}\partial^{i}\left(\delta\mathcal{E}\right)-\left(\frac{\partial \alpha}{\partial n}\right)_{\mathcal{E}}\frac{n_0T_0}{\left(\mathcal{E}_0+P_0\right)}\kappa T_{0}\partial^{i}\left(\delta n\right)-\kappa T_{0}\alpha_{0}\partial^{i}\Pi-\kappa T_{0}\alpha_{1}\partial_{j}\pi^{ij}&=0
\end{align}
\end{strip}

To study the dispersion relations of the hydrodynamic modes, we transform the linearized differential equations into momentum space by, 
\begin{equation}
    \delta \Psi(t,z)=\int \frac{d\omega \ dk}{\left(2\pi\right)^{2}}\delta\tilde{\Psi}\left(\omega,k\right)e^{-i\left(\omega t-kz\right)},
\end{equation}
where $\delta \Psi$ represents the state vector of perturbations. Applying this transformation to the linearized system yields the following coupled algebraic equations,

\begin{align}
    -i\omega \delta \mathcal{E}+ik \mathbb{W}_{0} \delta u^{z}&=0\\
    ik\left(c_{\mathcal{E}}^{2}\delta \mathcal{E}+ c_{n}^{2} \delta n + \Pi+\pi^{zz}\right)-i \omega \mathbb{W}_{0} \delta u^{z}&=0\\
    -i \omega \delta n + i k n_{0} \delta u^{z}-ik  \left(\frac{n_{0}}{\mathbb{W}_{0}}\right)q^{z}&=0\\
    ik\left(\zeta/3\right)\left(\delta u^{z}-\alpha_{0}q^{z}\right)+\left[1-i\omega \left(\frac{\zeta}{3}\right)\beta_{0}\right]\Pi&=0\\
    \left(1-2 i \eta \beta_{2} \omega\right)\pi^{zz}+\frac{4}{3}i \eta k \left(\delta u^{z}-\alpha_{1}q^{z}\right)&=0\\\nonumber{}
    i k \kappa T_{0}\left(\chi_{\mathcal{E}}\delta \mathcal{E}+\chi_{n}\delta n+\alpha_{0} \Pi+\alpha_{1}\pi^{zz}\right)&\\ 
    -\left(1-i \omega \kappa T_{0}\beta_{1}\right)q^{z}&=0,
\end{align}
where $c_{\mathcal{E}}^{2}=\left(\partial P/\partial \mathcal{E}\right)_{n}$, $c_{n}^{2}=\left(\partial P/\partial n\right)_{\mathcal{E}}$, $\mathbb{W}_{0}=\left(\mathcal{E}_{0}+P_{0}\right)$ and the thermodynamic susceptibilities are given by, $\chi_{\mathcal{E}}=\frac{n_{0}T_{0}}{\mathbb{W}_{0}}\left(\frac{\partial \alpha}{\partial \mathcal{E}}\right)_{n}$ and $\chi_{n}=\frac{n_{0}T_{0}}{\mathbb{W}_{0}}\left(\frac{\partial \alpha}{\partial n}\right)_{\mathcal{E}}$.

The above set of six equations can be represented as
\begin{equation}
    \mathcal{M}\left(\omega,k\right)\Psi=0,
\end{equation}
where, $\Psi=\left(\delta \mathcal{E}, \delta n, \delta u^{z}, \Pi, \pi^{zz}, \delta q^{z}\right)^{T}$ and 

\begin{strip}
\begin{equation}
\mathcal{M}(\omega, k) =\begin{bmatrix}
-i\omega & 0 & ik\mathbb{W}_{0} & 0 & 0 & 0 \\
0 & -i\omega & ikn_{0} & 0 & 0 & -ik\frac{n_{0}}{\mathbb{W}_{0}} \\
ikc_{\mathcal{E}}^{2} & ikc_{n}^{2} & -i\omega \mathbb{W}_{0} & ik & ik & 0 \\
0 & 0 & ik\frac{\zeta}{3} & 1-i\omega\frac{\zeta}{3}\beta_{0} & 0 & -ik\frac{\zeta}{3}\alpha_{0} \\
0 & 0 & ik\frac{4\eta}{3} & 0 & 1-2i\omega\eta\beta_{2} & -ik\frac{4\eta}{3}\alpha_{1} \\
-ik\kappa T_{0}\chi_{\mathcal{E}} & -ik\kappa T_{0}\chi_{n} & 0 & -ik\kappa T_{0}\alpha_{0} & -ik\kappa T_{0}\alpha_{1} & 1-i\omega\kappa T_{0}\beta_{1}
\end{bmatrix}
\end{equation}
\end{strip}

\section{Acoustic Attenuation and Dissipation Scales}

\subsection{Determination of The Threshold wavelength}

Our present analysis is restricted to the long-wavelength hydrodynamic regime, in which the sound wavelength is large compared to the correlation length, $k \xi << 1$. In this regime, the critical fluctuation may be treated as local equilibrium degrees of freedom and their effects are encoded in the local thermodynamic and transport coefficients entering through the acoustic dispersion relation. Although the correlation length $\left(\xi \right)$ is enhanced near the CEP, it remains finite in our present parameterized EoS framework. Hence, for sound modes with wavelengths comparable to the transverse size of the fireball, the long-wavelength condition can be satisfied over much of the system's evolution. In long wavelength approximation the sound dispersion relations take the following form: 

\begin{equation}
    \omega_{\pm}\left(k\right)=\pm c_{s}k-i\Gamma_{s}k^{2}+\mathcal{O}\left(k^{3}\right), \label{eqn48}
\end{equation}
where $c_{s}$ is the speed of sound and $\Gamma_{s}$ is the sound attenuation coefficient.
The sound attenuation coefficient is given by
\begin{equation}
    \Gamma_{s}=\frac{4\eta+\zeta}{6\mathbb{W}_{0}}+\frac{3\kappa n_{0}T_{0}c_{n}^{2}\left(\chi_{n}n_{0}+\chi_{\mathcal{E}}\mathbb{W}_{0}\right)}{6\mathbb{W}_{0}\left(c_{n}^{2}n_{0}+c_{\mathcal{E}}^{2}\mathbb{W}_{0}\right)},
\end{equation}

\begin{equation}
    \text{and}\ c_{s}^{2}=c_{\mathcal{E}}^{2}+c_{n}^{2}\left(\frac{n_{0}}{\mathbb{W}_{0}}\right),
\end{equation}

 \begin{equation*}
     \text{where,}\ c_{\mathcal{E}}^{2}=\frac{\left(\frac{\partial P}{\partial T}\right)_{\mu_B}\left(\frac{\partial n}{\partial \mu_{B}} \right)_{T}-\left(\frac{\partial P}{\partial \mu_{B}}\right)_{T}\left(\frac{\partial n}{\partial T}\right)_{\mu_{B}}}{\left(\frac{\partial \mathcal{E}}{\partial T}\right)_{\mu_{B}}\left(\frac{\partial \mathcal{E}}{\partial \mu_{B}}\right)_{T}-\left(\frac{\partial \mathcal{E}}{\partial \mu_{B}}\right)_{T}\left(\frac{\partial n}{\partial T}\right)_{\mu_{B}}},
 \end{equation*}

 \begin{equation*}
     \text{and},\ c_{n}^{2}=\frac{\left(\frac{\partial P}{\partial \mu_{B}}\right)_{T}\left(\frac{\partial \mathcal{E}}{\partial T}\right)_{\mu_{B}}-\left(\frac{\partial P}{\partial n }\right)_{\mu_{B}}\left(\frac{\partial \mathcal{E}}{\partial \mu_{B}}\right)_{T}}{\left(\frac{\partial n}{\partial \mu_{B}}\right)_{T}\left(\frac{\partial \mathcal{E}}{\partial T}\right)_{\mu_{B}}-\left(\frac{\partial n}{\partial T}\right)_{\mu_B}\left(\frac{\partial \mathcal{E}}{\partial \mu_{B}}\right)_{T}}.
 \end{equation*}

Now, from the sound wave dispersion relation we can obtain the threshold value of $k$ i.e. $k_{\text{th}}$, given by the following relation \cite{Liao:2009gb}:

\begin{equation}
    \left. \frac{\Im \left[\omega\left(k\right)\right]}{\Re \left[\omega\left(k\right)\right]}\right|_{k=k_{th}}=1, 
\end{equation}

Any sound wave with the wave vector beyond this limit gets attenuated in the fluid. From equation (\ref{eqn48}), we get $k_{th}=c_{s}/\Gamma_{s}$. The associated threshold wavelength is given by, 
\begin{equation}
    \lambda_{\text{th}}=\frac{2\pi\Gamma_{s}}{c_{s}}.
\end{equation}
This quantity provides a macroscopic estimate of the length scale below which sound perturbations are strongly attenuated.

\subsection{Linearized acoustic response to anisotropic flow}

Because anisotropic flow is driven by pressure gradients, the impact of
dissipative effects can be assessed through the attenuation of sound waves.
In the linear-response regime, the flow harmonics are related to the
corresponding initial spatial eccentricities by
\begin{equation}
v_n=\kappa_n\varepsilon_n,
\end{equation}
where $\varepsilon_n$ is the magnitude of the initial eccentricity vector and
$\kappa_n$ is the hydrodynamic response coefficient. This linear relation is
a good approximation for elliptic and triangular flow in a narrow centrality
interval, although nonlinear mode-coupled contributions become important for
higher harmonics and in more peripheral collisions.

Motivated by the acoustic scaling of anisotropic flow
\cite{Lacey:2013is,Lacey:2013eia}, we parametrize the dissipative
modification of the linear response as
\begin{equation}
v_n^{\mathrm{linear}}
=
\kappa_n^{(0)}\varepsilon_n
\exp\left[-\beta_{\mathrm{eff}}\frac{n^2}{\bar R}\right],
\end{equation}
where $\kappa_n^{(0)}$ is a non-dissipative reference response coefficient,
$\bar R$ is an effective transverse size of the system, and
$\beta_{\mathrm{eff}}$ is an effective acoustic-damping coefficient with
dimensions of length. The coefficient $\beta_{\mathrm{eff}}$ is not uniquely
determined by the homogeneous linear dispersion relation. Rather, it
parametrizes the integrated damping of sound perturbations in an expanding,
finite-lived, and spatially inhomogeneous medium. Its dependence on the
thermodynamic variables is motivated by the behavior of the sound attenuation
coefficient $\Gamma_s$.

If the effective wave number of a geometric perturbation is associated with
the harmonic number through $k\sim n/\bar R$, the plane-wave damping factor
over a characteristic evolution time $t$ is estimated as
\begin{equation}
S_n\sim \exp\left[-\frac{n^2\Gamma_s t}{\bar R^2}\right].
\end{equation}
For a given value of
$\beta_{\mathrm{eff}}/\bar R$, triangular flow is more strongly suppressed
than elliptic flow. However, it should not be interpreted as an exact
derivation of $v_{n}$, because the mapping from a static
plane-wave perturbation to the final anisotropic flow depends on the
expansion dynamics, which is beyond the scope of the linear-response
treatment.

The initial eccentricity vector can be evaluated from the transverse
reduced-thickness profile $T_R(x,y)$ as
\begin{equation}
\mathcal E_n=\varepsilon_ne^{in\Psi_n}= -\frac{\int d^2x_\perp\,r^ne^{in\phi}T_R(x,y)}{\int d^2x_\perp\,r^nT_R(x,y)}.
\end{equation}
The eccentricity magnitude is $\varepsilon_n=|\mathcal{E_{n}}|$. The effective
transverse size is characterized by the rms radius,
\begin{equation}
\bar R^2=
\frac{
\int d^2x_\perp\,r^2T_R(x,y)
}{
\int d^2x_\perp\,T_R(x,y)
}.
\end{equation}

To illustrate the magnitude of the linear acoustic response, we generate
mid-central Au+Au initial conditions at beam energies relevant to the RHIC
Beam Energy Scan using the
TRENTo initial-condition model
\cite{Moreland:2014oya}. For each event ensemble, we compute
the rms eccentricities $\varepsilon_2\{2\}$ and $\varepsilon_3\{2\}$, as well
as the effective transverse radius $\bar R$. Also we adopt representative linear
response coefficients $\kappa_2^{(0)}$ and $\kappa_3^{(0)}$ consistent with
hydrodynamic analyses of mid-central heavy-ion collisions at these energies.

\section{Results and discussion}
\subsection{Equation of state and critical sound attenuation}

\begin{figure}[!h]
    \centering
    \includegraphics[width=0.5\textwidth]{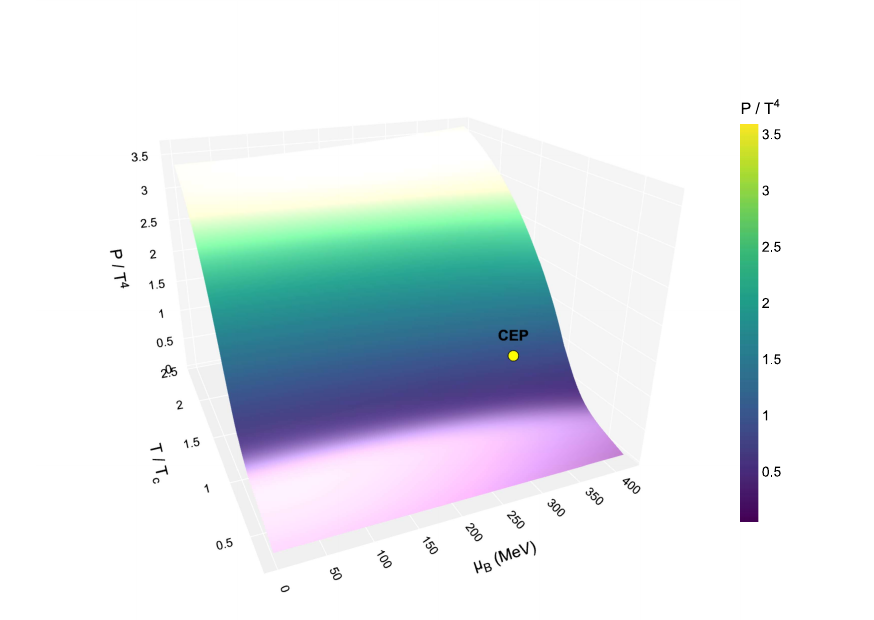}
    \caption{Variation of the normalized thermodynamic pressure $P/T^{4}$ with respect to normalized temperature $T/T_{c}$ and baryon chemical potential $\mu_{B}$}
    \label{fig_Pressure}
\end{figure}

\begin{figure}[!h]
    \centering
    \includegraphics[width=0.5
    \textwidth]{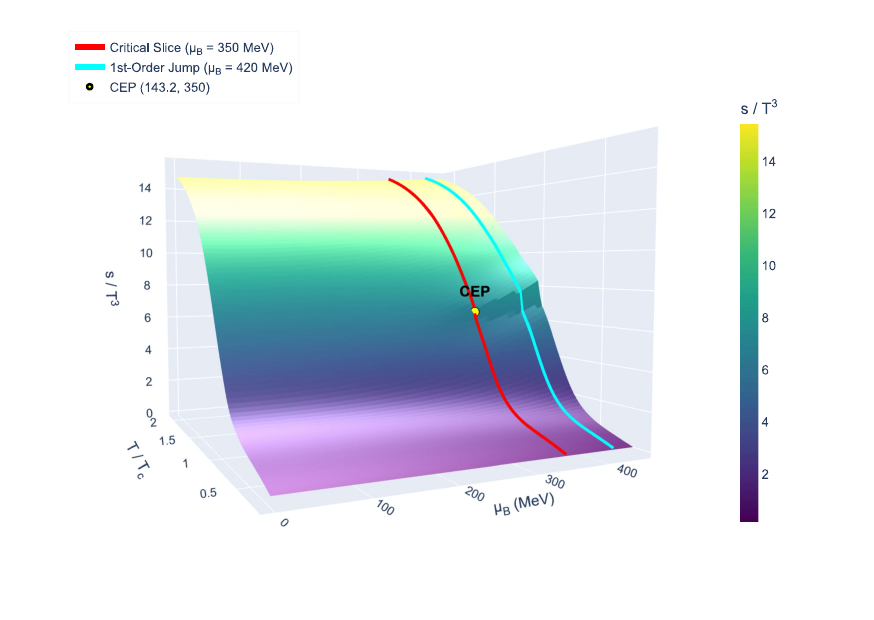}
    \caption{Variation of normalized entropy $S/T^{3}$ w.r.t $T/T_{c}$ and $\mu_{B}$. The red line $\left( \text{at}~\mu_{B}=350~\text{MeV}\right)$ in the diagram indicates a transition line between a crossover and $1^{st}$ order phase transition. The CEP is located at $\left(T_{c},\mu_{Bc}\right)=\left(143.2,350\right)~\text{MeV}.$} 
    \label{fig_entropy}
\end{figure}

We begin our discussion by examining the parameterized equation of state. Fig. \ref{fig_Pressure} depicts the variation of pressure $P$ with respect to temperature $T$ and chemical potential $\mu_{B}$. In the vicinity of the critical point, the pressure remains continuous, however, the entropy shows a rapid variation (fig \ref{fig_entropy}). In fig. \ref{fig_entropy}, for $\mu_{B}< \mu_{Bc}$, there is no significant change in the entropy, implying a smooth crossover; but for $\mu_{B}> \mu_{Bc}$, a distinct jump is observed, indicating first order phase transition.

\begin{figure}[!h]
    \centering
    \includegraphics[width=0.5
    \textwidth]{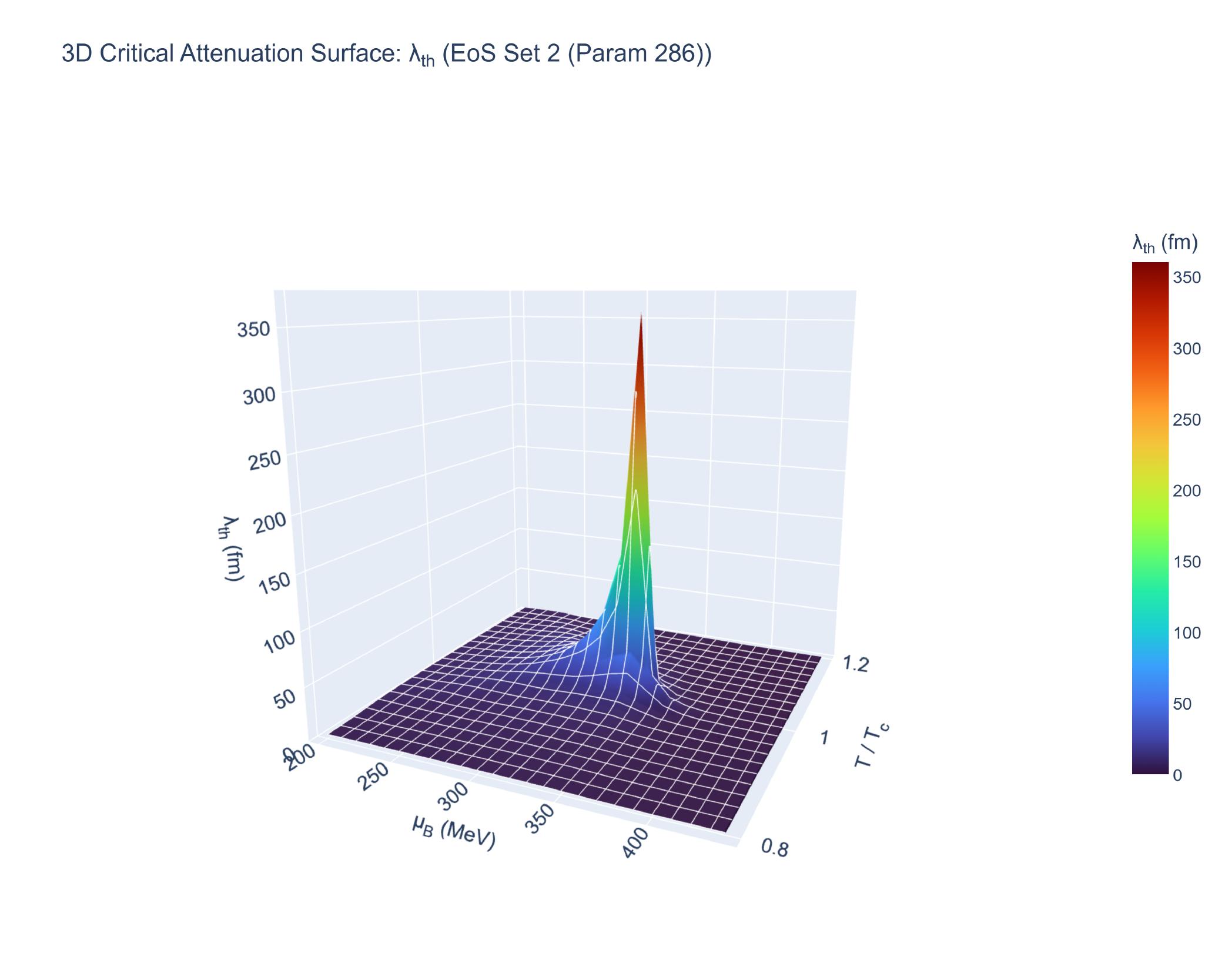}
    \caption{Variation of threshold wavelength $\lambda_{\text{th}}$ with respect to $T/T_{c}$ and $\mu_{B}$. The non-universal parameters used for the EOS are $w=2$ and $\rho=1$.}
    \label{fig_lambda}
\end{figure}

Such critical thermodynamics  directly impacts the acoustic propagation within the medium. Fig. \ref{fig_lambda} shows the threshold wavelength $\lambda_{\text{th}}$ in the $T-\mu_{B}$ plane.
The magnitude and width of this threshold wavelength is governed by various thermodynamic quantities, transport coefficients and relaxation times.

In this present calculation, we have considered $\eta/s=1/4\pi$, $\kappa/s=0.05$ and $\zeta/s\propto\xi^{3}$. Near the critical end point, $\lambda_{\text{th}}$ develops a pronounced peak and decreases gradually as the system moves away from the critical region. 

For a sound perturbation of wavelength $\lambda$, attenuation takes place wherever $\lambda\lesssim \lambda_{\text{th}}$. Near the CEP, $\lambda_{\text{th}}$ becomes very large as compared  to the characteristic transverse radius of the QGP fireball, $\bar{R}$. Thus perturbations of length scales comparable to or smaller than the system size can get strongly damped near CEP. This indicates that the critical enhancement of sound attenuation can substantially reduce the ability of the medium to sustain propagating sound mode on length scales relevant to heavy ion collisions.

\subsection{Parameter sensitivity of the EoS}

The mapping of the $3D$ Ising model onto the QCD  phase diagram cannot be fixed by using the first principle, so we systematically vary the non-universal parameters $\rho$ and $w$. Variation of these two parameters modifies the thermodynamic structure of the critical region and consequently altering the magnitude and spatial extent of 
$\lambda_{th}$.

\begin{figure}[!h]
    \centering
    \includegraphics[width=0.5\textwidth]{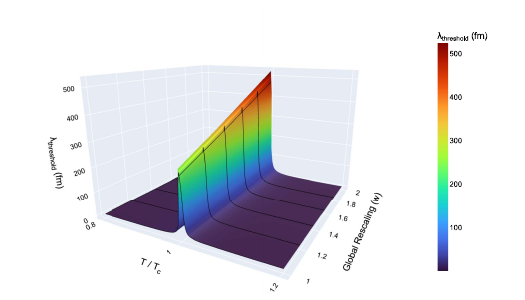}
    \caption{Variation of the global rescaling parameter $w$ with temperature $T/T_{c}$ and baryon chemical potential $\mu_{B}$ Here $\rho=1$ and $1.00<w<2.00$}
    \label{fig_rho_fix_w_variable_3D}
\end{figure}

\begin{figure*}[!t]
  \centering
  \includegraphics[width=15 cm, height=5 cm]{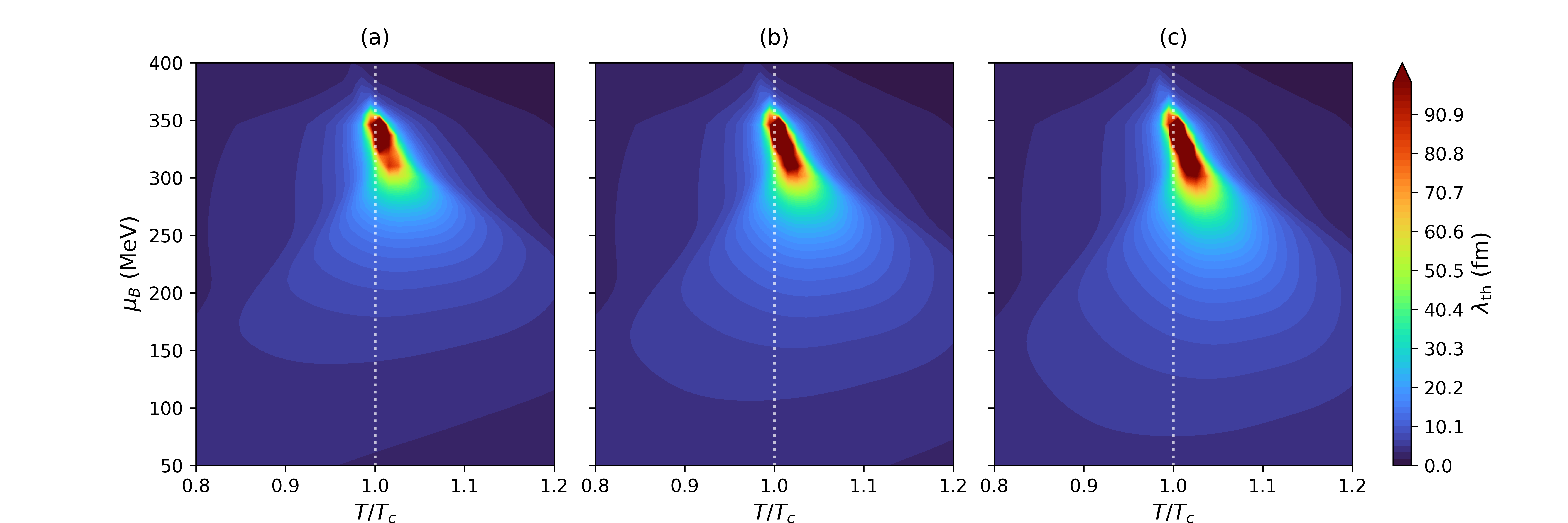}
  \caption{Variation of the spatial extent of the threshold wavelength $\lambda_{\text{th}}$ with respect to $\mu_{B}$ and $T/T_{c}$. In fig (a) $w=1.5$ and $\rho=1.50$, in fig (b) $w=1.5$ and $\rho=1.75$ and in fig (c) $w=1.5$ and $\rho=2.00$.}
  \label{muB_vs_T_w-fixed_2D}
\end{figure*}

Keeping the relative rescaling parameter $\rho$ fixed and varying the global rescaling parameter $w$ modifies the peak magnitude of $\lambda_{th}$ (fig \ref{fig_rho_fix_w_variable_3D}).The corresponding spatial extent of the threshold wavelength is shown in fig. \ref{muB_vs_T_w-fixed_2D}.

\begin{figure*}[!t]
  \centering
  \includegraphics[width=15 cm, height=5 cm]{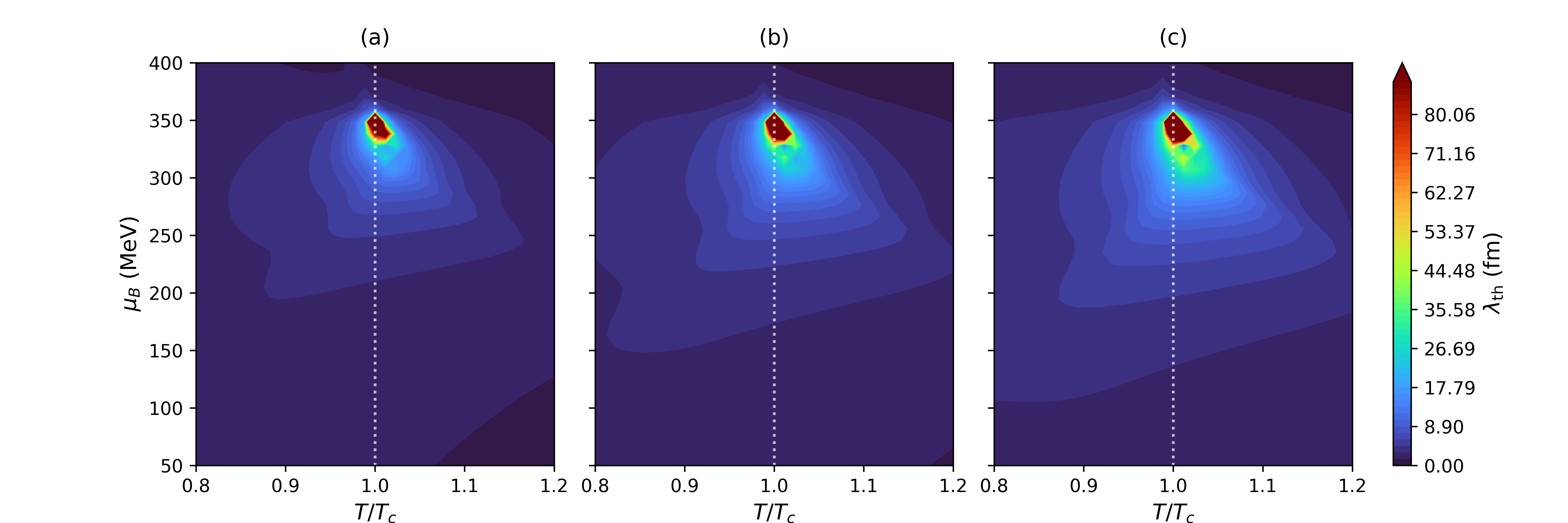}
  \caption{Variation of the spatial extent of the threshold wavelength $\lambda_{\text{th}}$ with respect to $\mu_{B}$ and $T/T_{c}$. In fig (a) $\rho=1$ and $w=1.00$, in fig (b) $\rho=1$ and $w=1.25$ and in fig (c) $\rho=1$ and $w=1.50$.}
  \label{muB_vs_T_rho-fixed_2D}
\end{figure*}

\begin{figure}[!h]
    \centering
    \includegraphics[width=0.5\textwidth]{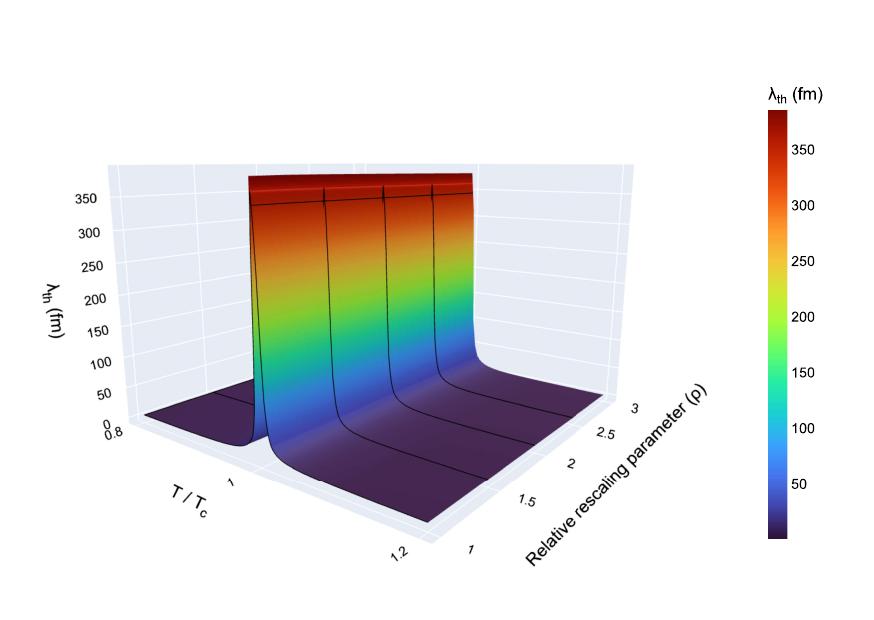}
    \caption{Variation of the relative rescaling parameter $\rho$ with temperature $T/T_{c}$ and baryon chemical potential $\mu_{B}$. Here $w=1.5$ and $1.00<\rho<3.00$}
    \label{fig_rho_fix_rho_variable_3D} 
\end{figure}

On the other hand, holding $w$ fixed and varying $\rho$ changes the spatial extent of $\lambda_{\text{th}}$ (fig. \ref{muB_vs_T_w-fixed_2D}), but the peak magnitude of $\lambda_{th}$ remains constant (fig. \ref{fig_rho_fix_rho_variable_3D}).

Therefore, the sound attenuation effects are parameter dependent, although the qualitative enhancement is a common feature of the EoS parameter range.

\subsection{Linear harmonic response}

We have mapped the critical attenuation profile onto the linear response of anisotropic flow to estimate how the attenuation of sound perturbations modify the conversion efficiency of initial spatial eccentricities into final-state momentum anisotropies. The estimated elliptic and triangular flow responses, $v_2$ and $v_3$, are obtained from the acoustic-response parametrization
\begin{equation}
v_n^{linear}(T,\mu_B) = \kappa_n^{(0)}\varepsilon_n \exp\left[-\beta_{\mathrm{eff}}(T,\mu_B)\frac{n^2}{\bar R}\right], ~n=2,3.
\end{equation}
Here, $\varepsilon_n$ and $\bar R$ are determined from representative initial transverse profiles, while $\beta_{\mathrm{eff}}(T,\mu_B)$ is an integrated damping parameter motivated by the sound attenuation coefficient obtained from the dispersion analysis.

\begin{figure}[!h]
    \centering
    \includegraphics[width=0.45\textwidth]{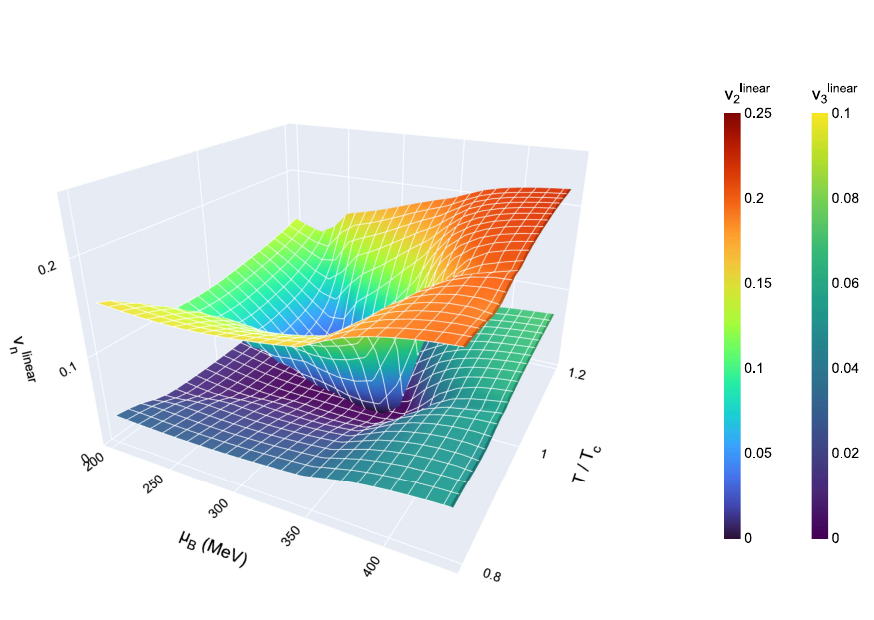}
    \caption{Variation of linearized flow harmonics $v_{n}^{\text{linear}}$ with respect to $T/T_{c}$ and $\mu_{B}$.}
    \label{fig_vn}
\end{figure}

\begin{figure}[!h]
	\centering
	\includegraphics[width=0.45\textwidth]{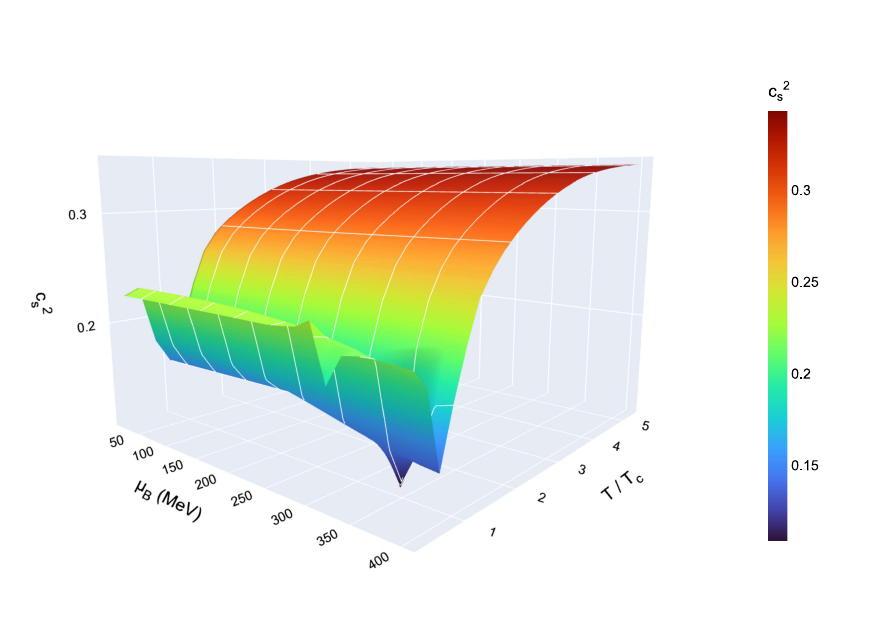}
	\caption{Variation of $c_{s}^{2}$ with respect to $T/T_{c}$ and $\mu_{B}$. The $c_{s}^{2}$ takes a sharp dip at $\left(T_{c},\mu_{Bc}\right)$.}
	\label{fig_cs2}
\end{figure}

The estimated flow responses reflects an asymmetry around $T_{c}$. Above $T_{c}$, the system is dominated by deconfined quarks and gluons. Below $T_{c}$, the system evolves towards a hadronic resonance gas (HRG) regime populated by massive hadrons, leading to a softer EoS and a reduced speed of sound (fig. \ref{fig_cs2}). Fig. (\ref{fig_vn}) shows that the linear response of triangular flow $v_{3}$ is more strongly suppressed the linear response of the elliptical flow $v_{2}$. This present estimation should be considered as a phenomenological measure of the impact of the critical sound attenuation on the linear conversion efficiency $v_{n}/\varepsilon_{n}$, rather than as direct predictions for experimentally measured flow coefficients.

\section{Summary and Conclusions}

We have investigated the linearized sound-channel response of a finite-baryon-density Israel--Stewart hydrodynamic system coupled to a parameterized equation of state containing a critical end point. Sound attenuation is characterized by the threshold wavelength $\lambda_{\mathrm{th}}$, which defines the length scale below which sound perturbations are strongly damped. This quantity provides a macroscopic measure of the demarcation between sound propagation and dissipation in baryon-rich QCD matter.

In the thermodynamic limit, $\lambda_{\mathrm{th}}$ diverges at the CEP \cite{Hasanujjaman2020}, implying that sound modes on all relevant length scales are dissipated. In the present parameterized treatment, however, $\lambda_{\mathrm{th}}$ remains finite because of the adopted parametrization of the critical thermodynamics and the finite correlation length. By varying the nonuniversal parameters that control the critical region within the BEST equation-of-state framework, we examined how the magnitude and width of the peak in $\lambda_{\mathrm{th}}$ change near the CEP-like region. The resulting peak can exceed the characteristic transverse size of the QGP fireball by a substantial margin, indicating strong absorption of sound modes on fireball-scale wavelengths.

The enhanced sound absorption has two related phenomenological consequences. First, it reduces the ability of the medium to support Mach-cone-like correlations. In the idealized thermodynamic limit, the divergence of $\lambda_{\mathrm{th}}$ at the CEP would imply a vanishing Mach angle and the disappearance of the Mach-cone signal. In the present finite-size treatment, one expects a strong suppression rather than a strict absence of the Mach-cone response. Second, the linear conversion of initial spatial eccentricities into final anisotropic flow is reduced near the CEP. The estimated linear elliptic and triangular flow responses are strongly attenuated in this region, with a stronger suppression of $v_3$ than $v_2$.

In the Beam Energy Scan program, varying the collision energy $\sqrt{S_{NN}}$ causes the system to follow diffferent trajectories through the $T-\mu_{B}$ plane, along which $\lambda_{\text{th}}$ is evaluated. The beam energy at which $\lambda_{\text{th}}$ is maximum, identifies the trajectory that samples the critical end point most closely. The associated suppression of Mach cone like correlations and anisotrpic flow is a collective consequence of this critical traversal. Thus a non-monotonic bean energy scan dependence of the $\lambda_{\text{th}}$ along with the reduced Mach cone like correlation and stronger suppression of $v_{3}$ than $v_{2}$, would provide a phenomenological signature that the system trajectory has passed close to the CEP of the chosen EoS.

Several limitations of the present analysis should be emphasized. The finite peak of $\lambda_{\mathrm{th}}$ reflects the adopted parametrization of the critical thermodynamics and the finite correlation length, rather than a first-principles calculation of the dynamical limitation imposed by critical slowing down. Near the CEP, the relaxation of long-wavelength critical fluctuations becomes parametrically slow; a quantitative treatment of this effect requires an extended dynamical framework, such as Hydro+ \cite{Stephanov:2017ghc}, which is beyond the scope of the present work. Moreover, the estimated anisotropic-flow response is linearized and phenomenological, whereas Mach-cone formation and its survival until freeze-out depend on nonlinear expansion dynamics, jet energy loss, viscosity, initial-state fluctuations and noncritical transport properties. A quantitative identification of the QCD critical point would therefore require a self-consistent dynamical treatment of critical fluctuations, finite-time evolution, and the full space-time evolution of the collision system.

Nevertheless, the present results provide a useful theoretical reference for interpreting anisotropic-flow measurements and for assessing acoustic signatures of a CEP in baryon-rich QCD matter. They may be relevant for the RHIC-E at Beam Energy Scan program and for future high-baryon-density experiments, including CBM at FAIR.

\begin{acknowledgements}
AP, MH and PKH are grateful to the Inter-University Center for Astronomy and Astrophysics (IUCAA), Pune, India for providing Associateship Programme under which a part of this work was carried out.
\end{acknowledgements}

\section*{Data Availability Statement}
The data used in this work to generate the equation of state, as well as the initial condition data from the heavy-ion collisions, will be made available by the corresponding author upon reasonable request.

\section*{Code Availability Statement}
The custom numerical routines used to compute the hydrodynamic dispersion relations and flow estimates are available from the corresponding author upon reasonable request. 

\section*{Conflict of interest}
The authors declare that they have no known financial interest or personal relationship that could have appeared to influence the work reported in this paper.

\bibliographystyle{spphys}
\bibliography{sorsamp} 

\providecommand{\noopsort}[1]{}\providecommand{\singleletter}[1]{#1}%
\begin{thebibliography}{10}
\providecommand{\url}[1]{{#1}}
\providecommand{\urlprefix}{URL }
\expandafter\ifx\csname urlstyle\endcsname\relax
  \providecommand{\doi}[1]{DOI \discretionary{}{}{}#1}\else
  \providecommand{\doi}{DOI \discretionary{}{}{}\begingroup
  \urlstyle{rm}\Url}\fi

\bibitem{wong1994}
C.Y. Wong, \emph{Introduction to high-energy heavy ion collisions} (World
  Scientific, 1994)

\bibitem{aoki2009}
Y.~Aoki, S.~Borsanyi, S.~Durr, Z.~Fodor, S.D. Katz, S.~Krieg, K.K. Szabo, JHEP
  \textbf{06}, 088 (2009)

\bibitem{borsanyi2010}
S.~Borsanyi, Z.~Fodor, C.~Hoelbling, S.D. Katz, S.~Krieg, C.~Ratti, K.K. Szabo,
  JHEP \textbf{09}, 073 (2010)

\bibitem{bazavov2012}
A.~Bazavov, et~al., Phys. Rev. D \textbf{85}, 054503 (2012)

\bibitem{Asakawa1989}
M.~Asakawa, K.~Yazaki, Nucl. Phys. A \textbf{504}, 668 (1989)

\bibitem{Halasz1998}
A.M. Halasz, A.D. Jackson, R.E. Shrock, M.A. Stephanov, J.J.M. Verbaarschot,
  Phys. Rev. D \textbf{58}, 096007 (1998)

\bibitem{Stephanov2004}
M.A. Stephanov, Prog. Theor. Phys. Suppl. \textbf{153}, 139 (2004)

\bibitem{Ghosh2001}
D.~Ghosh, A.~Deb, S.R. Sahoo, P.K. Haldar, Europhys. Lett. \textbf{56}(5), 639
  (2001)

\bibitem{Ghosh2004}
D.~Ghosh, A.~Deb, P.K. Haldar, S.R. Sahoo, D.~Maity, Europhys. Lett.
  \textbf{65}(3), 311 (2004)

\bibitem{Heinz2013}
U.~Heinz, R.~Snellings, Annual Review of Nuclear and Particle Science
  \textbf{63}(1), 123 (2013)

\bibitem{Gavai2003}
R.V. Gavai, S.~Gupta, Phys. Rev. D \textbf{68}(3), 034506 (2003)

\bibitem{Allton2005}
C.R. Allton, M.~Doring, S.~Ejiri, S.J. Hands, O.~Kaczmarek, F.~Karsch,
  E.~Laermann, K.~Redlich, Phys. Rev. D \textbf{71}(5), 054508 (2005)

\bibitem{Bellwied2015}
R.~Bellwied, S.~Borsanyi, Z.~Fodor, S.D. Katz, A.~Pasztor, C.~Ratti, K.K.
  Szabo, Phys. Rev. D \textbf{92}(11), 114505 (2015)

\bibitem{Borsanyi2010bp}
S.~Borsanyi, Z.~Fodor, C.~Hoelbling, S.D. Katz, S.~Krieg, C.~Ratti, K.K. Szabo,
  JHEP \textbf{09}, 073 (2010)

\bibitem{Bazavov2017}
A.~Bazavov, et~al., Phys. Rev. D \textbf{95}(5), 054504 (2017)

\bibitem{Bazavov2020}
A.~Bazavov, D.~Bollweg, H.T. Ding, P.~Enns, J.~Goswami, P.~Hegde, O.~Kaczmarek,
  F.~Karsch, R.~Larsen, et~al., Phys. Rev. D \textbf{101}(7), 074502 (2020)

\bibitem{Bellwied2015b}
R.~Bellwied, S.~Borsanyi, Z.~Fodor, J.~G{\"u}nther, S.D. Katz, C.~Ratti, K.K.
  Szabo, Phys. Lett. B \textbf{751}, 559 (2015)

\bibitem{Borsanyi2020}
S.~Borsanyi, Z.~Fodor, J.N. Guenther, R.~Kara, S.D. Katz, P.~Parotto,
  A.~Pasztor, C.~Ratti, K.K. Szabo, Phys. Rev. Lett. \textbf{125}(5), 052001
  (2020)

\bibitem{Borsanyi2018}
S.~Borsanyi, Z.~Fodor, J.N. Guenther, S.K. Katz, K.K. Szabo, A.~Pasztor,
  I.~Portillo, C.~Ratti, JHEP \textbf{10}, 205 (2018)

\bibitem{Parotto2020}
P.~Parotto, M.~Bluhm, D.~Mroczek, M.~Nahrgang, J.~Noronha-Hostler,
  K.~Rajagopal, C.~Ratti, T.~Sch{\"a}fer, M.~Stephanov, Phys. Rev. C
  \textbf{101}(3), 034901 (2020)

\bibitem{Karthein2021}
J.M. Karthein, D.~Mroczek, A.R. Nava~Acuna, J.~Noronha-Hostler, P.~Parotto,
  D.R.P. Price, C.~Ratti, Eur. Phys. J. Plus \textbf{136}(6), 621 (2021)

\bibitem{Hasanujjaman2020}
M.~Hasanujjaman, M.~Rahaman, A.~Bhattacharyya, J.e. Alam, Phys. Rev. C
  \textbf{102}(3), 034910 (2020)

\bibitem{Israel1979}
W.~Israel, J.M. Stewart, Annals of Physics \textbf{118}(2), 341 (1979)

\bibitem{Nonaka:2004kp}
C.~Nonaka, M.~Asakawa, Phys. Rev. C \textbf{71}, 044904 (2005)

\bibitem{Guida:1996ep}
R.~Guida, J.~Zinn-Justin, Nucl. Phys. B \textbf{489}, 626 (1997)

\bibitem{Schofield:1969zz}
P.~Schofield, J.D. Litster, J.T. Ho, Phys. Rev. Lett. \textbf{23}(19), 1098
  (1969)

\bibitem{Bluhm:2006CPOD}
M.~Bluhm, B.~K{\"a}mpfer, PoS \textbf{CPOD2006}, 004 (2006)

\bibitem{Parotto:2019qcd}
P.~Parotto, Nucl. Phys. A \textbf{982}, 183 (2019)

\bibitem{Landau1987}
L.D. Landau, E.M. Lifshitz, \emph{Fluid Mechanics}, 2nd edn.
\newblock Course of Theoretical Physics, Volume 6 (Pergamon Press, Oxford,
  1987)

\bibitem{Muronga:2006zx}
A.~Muronga, Phys. Rev. C \textbf{76}, 014909 (2007)

\bibitem{Liao:2009gb}
J.~Liao, V.~Koch, Phys. Rev. C \textbf{81}, 014902 (2010)

\bibitem{Lacey:2013is}
R.A. Lacey, A.~Taranenko, J.~Jia, D.~Reynolds, N.N. Ajitanand, J.M. Alexander,
  Y.~Gu, A.~Mwai, Phys. Rev. Lett. \textbf{112}, 082302 (2014)

\bibitem{Lacey:2013eia}
R.A. Lacey, D.~Reynolds, A.~Taranenko, N.N. Ajitanand, J.M. Alexander, F.H.
  Liu, et~al., J. Phys. G \textbf{43}(10), 10LT01 (2016)

\bibitem{Moreland:2014oya}
J.S. Moreland, J.E. Bernhard, S.A. Bass, Phys. Rev. C \textbf{92}, 011901
  (2015)

\bibitem{Stephanov:2017ghc}
M.~Stephanov, Y.~Yin, Phys. Rev. D \textbf{98}(3), 036006 (2018)

\end{thebibliography}

\end{document}